\documentclass[useAMS,usenatbib,onecolumn,usegraphicx]{mn2e}

\newcommand{\beq}{\begin{equation}}
\newcommand{\eeq}{\end{equation}}
\newcommand{\beqar}{\begin{eqnarray}}
\newcommand{\eeqar}{\end{eqnarray}}

\newcommand{\gsim}{\, \raisebox{-0.8ex}{$\stackrel{\textstyle >}{\sim}$ }}

\title[Alfv\'{e}n Polar Oscillations of Relativistic Stars]
{Alfv\'{e}n Polar Oscillations of Relativistic Stars}
\author[H.~Sotani \& K.~D. Kokkotas] 
{Hajime~Sotani$^1$\thanks{E-mail:sotani@astro.auth.gr} and
 Kostas~D.~Kokkotas$^{1,2}$\thanks{E-mail:kostas.kokkotas@uni-tuebingen.de}
\\
$^1$Theoretical Astrophysics, Eberhard-Karls University of T\"ubingen,
   72076, T\"ubingen, Germany \\
$^2$Department of Physics, Aristotle University of Thessaloniki,
  Thessaloniki 54124, Greece}

\begin{document}



\maketitle

\label{firstpage}

\begin{abstract}
 We study polar Alfv\'{e}n oscillations of relativistic stars endowed with a strong global poloidal dipole magnetic field. 
 Here we focus only on the axisymmetric oscillations which are studied by evolving numerically the two-dimensional perturbation equations.
Our study shows that the  spectrum of the polar
Alfv\'{e}n oscillations is discrete in contrast to the spectrum of  axial Alfv\'{e}n oscillations
which is continuous.
We also show that the typical fluid modes, such as the $f$
and $p$ modes, are not significantly affected by the presence of the strong magnetic field.
\end{abstract}

\section{Introduction}
\label{sec:Intro}

The soft gamma repeaters (SGRs) are the objects radiating sporadic X-ray and gamma-ray
bursts, whose typical luminosities are around $10^{41}$ erg/s. Rarely the SGRs were associated with
the emission of stronger gamma-rays with peak luminosities of about $10^{44}\sim 10^{46}$ erg/s.
These phenomena are typically referred as ``giant flares". Up to now
at least three giant flares have been detected in three different SGRs,
which are : the SGR 0526--66 in 1979, the SGR 1900+14 in 1998 and the SGR 1806--20 in 2004.
During these giant  flares in SGRs  we observe an initial strong
sharp burst followed by a  decaying tail which may last for hundreds of seconds. The timing analysis of
the decaying tail of the latest two giant flares revealed the existence of
quasi periodic oscillations (QPOs), whose specific frequencies are approximately
18, 26, 30, 92, 150, 625 and 1840 Hz for SGR 1806--20, and 28, 53, 84 and 155 Hz for SGR 1900+14
(see \cite{WS2006} for a review). The most promising mechanism to produce these QPOs are
are the magnetar models. Magnetars are thought to be  neutron stars with strong magnetic fields, 
$B\gsim 5\times 10^{13}$ Gauss \citep{DT1992}. According to this scenario, starquakes may be
driven during the release of the accumulated magnetic stress in the crust region.
The excited seismic oscillations can be decomposed into two types, i.e. the axial-type
and the polar-type oscillations. Since the axial-type oscillations do not involve 
density variations, it is thought that it will be easier to excite them instead of the polar-type ones. In this case, the magnetic field globally and/or the shear modulus in the crust,
act as the necessary restoring force for the axial-type oscillations of the non-rotating
magnetars.

For the last few years the attempts to explain the above QPO frequencies in SGRs
by using the crust torsional oscillations have been based both in Newtonian (e.g., \cite{Lee2007})
and in general relativistic dynamics (e.g., \cite{SA2007}; \cite{SKS2007}). These attempts
were partially successful, i.e., some of the observed frequencies are in good agreement with the
frequencies of torsional oscillations. But very soon it has been realized the difficulty
to explain all the observed frequencies in SGRs by using only the  torsional oscillations of the crust (magnetized or not).
In an attempt to explain the QPOs without making use of the crust torsional oscillations,
\cite{Levin2006} using a toy model suggested that the torsional Alfv\'{e}n oscillations could not
form a discrete spectrum but instead they should form a continuum. He strongly pointed out
the importance for the crust-core coupling and soon later \cite{GSA2006}
came with a similar suggestion i.e. that the QPO spectrum may be explained by considering  
the torsional Alfv\'{e}n oscillations of the magnetar's core.
About a year later, this suggestion has been verified with a more elaborate model by \cite{Levin2007},  while next year   \citet{SKS2008} working on a more realistic model showed that the spectrum of the torsional Alfv\'{e}n oscillations of relativistic magnetars forms a continuum. Moreover, they demonstrated  that there exist two distinct families
as the so called ``lower" and ``upper" QPOs. Finally, they figured out that the observed QPO frequencies in SGRs
(at least the lower observed frequencies) can be explained by using the frequencies of torsional
Alfv\'{e}n oscillations as well as the crust torsional oscillations.

This last result about the  torsional Alfv\'{e}n oscillations, has been studied in more detail
in two recent papers by \cite{CBK2009} and \cite{DSF2009}.
One could actually argue that the features of torsional Alfv\'{e}n oscillations are quite well understood.
However, the spectrum of the polar Alfv\'{e}n  oscillations of magnetars is quite unexplored.
So far, there are only a few studies of polar oscillations of the magnetized
stars, where for example, they consider only the fluid modes for Ap stars \citep{ST1993},
the oscillations in a spherical shell for Newtonian stellar models with an incompressible
fluid \citep{RR2003}, and the oscillations in the crust region of Newtonian models \citep{Lee2007}.

From the previous discussion one can realize that the polar part of the spectrum of relativistic magnetized stars has not yet been studied in detail while it is still  unknown whether spectrum of the polar Alfv\'{e}n oscillations is discrete
or continuous. Moreover, some of the higher QPO frequencies observed may be related to polar and not to axial Alfv\'{e}n oscillations. Finally, polar oscillations are related to the emission of gravitational waves and the ultimate result will be  the estimation of the amount of the energy released in the form of gravitational waves during the giant flares \citep{Abbott2007}. 

In an attempt to shed some light to the various features of polar
Alfv\'{e}n oscillations,  we performed two-dimensional numerical evolutions of the perturbation equation 
of relativistic magnetar models endowed with a global poloidal dipole magnetic field. It should be mentioned here that in the present study we omit the effects due to the presence of a crust and we focus only on the axisymmetric polar oscillations of the core. The omission of the crust simplifies considerably the boundary conditions of the problem  without affecting significantly the actual frequencies of the oscillation modes. As a future project we can foresee the inclusion of the crust which induces additional families of shear and interfacial modes (e.g., \citet{VKS2008}) which  can be used to explain the observed QPO frequencies. In other words, this article is the first step towards a detailed study of the  oscillation spectrum of magnetized relativistic stars. 

This paper is structured as follows. In the next section we describe the equilibrium configuration
adopted in this paper. In section \ref{sec:III} we drive the basic perturbation equations with
the appropriate boundary conditions. The numerical results are shown in section \ref{sec:IV},
and finally we give a conclusion in section \ref{sec:V}.
Unless otherwise noted, we adopt units of $c=G=1$, where $c$ and $G$ denote the speed of light
and the gravitational constant, respectively, while the metric signature is $(-,+,+,+)$.

\section{Equilibrium configuration}
\label{sec:II}

We assume that the background models of the strongly magnetized stars that we will study,
are  spherically symmetric and non-rotating. This choice is justified since all known magnetars rotate very slowly,
i.e., periods of a few seconds while   
it is true that even for magnetars with strong magnetic fields such as $B\gsim 10^{15}$ Gauss,
the deformation due to the magnetic pressure is not significant (see for example \cite{Colaiuda2008};
\cite{Haskell2008})
\footnote{In this paper, for simplicity, we will not take into account the  deformation of the star
due to the presence of the strong magnetic field. These type of deformations will potentially affect, quantitatively,  the oscillation spectrum, although one should not expect any important qualitative effect.}.
This is expected since even for the case of magnetars the magnetic energy is much smaller than the gravitational binding energy,
i.e., ${\cal E}_m/{\cal E}_g \sim 10^{-4}(B/10^{16}$ Gauss)$^2$, where ${\cal E}_m$
and ${\cal E}_g$ are the magnetic and gravitational binding energies, respectively.
In other words, we consider a static spherically symmetric neutron star,
which is a solution of the Tolman-Oppenheimer-Volkov
(TOV) equations. In this case the metric is defined as
\begin{equation}
 ds^2 = -e^{2\Phi(r)} dt^2 + e^{2\Lambda(r)} dr^2 + r^2 \left(d\theta^2 + \sin^2\theta d\phi^2\right),
\end{equation}
where $e^{-2\Lambda} = 1-2m(r)/r$ and the fluid four velocity is $u^\mu = (e^{-\Phi},0,0,0)$.
On this stellar model we superimpose a dipole magnetic field.
We assume that the star consists of a perfect fluid and in addition we make the assumption of the ``the ideal MHD approximation''.
This implies that a comoving observer cannot feel the presence of an electric field.
Then the stress-energy tensor is given by
\begin{equation}
 T^{\mu\nu} = (\epsilon + p)u^\mu u^\nu + pg^{\mu\nu}
            + H^\alpha H_\alpha \left(u^\mu u^\nu + \frac{1}{2}g^{\mu\nu}\right) - H^\mu H^\nu,
\end{equation}
where $\epsilon$, $p$, and $H^\mu$ are the energy density, the pressure, and the normalized magnetic
field defined as $H^\mu = B^\mu / \sqrt{4\pi}$, respectively.
In particular, in this paper we adopt the pure poloidal dipole magnetic field, while the effect due to
the toroidal magnetic fields will be left for future studies. The pure poloidal magnetic field
is given as
\begin{equation}
 (H_r,H_\theta,H_\phi)
    = \left(\frac{e^{\Lambda}a_1(r)}{\sqrt{\pi}r^2}\cos\theta,
       -\frac{e^{-\Lambda}\partial_r a_1(r)}{\sqrt{4\pi}}\sin\theta,0\right),
    \label{DMF}
\end{equation}
where $a_1(r)$ is a radial component of the electromagnetic four-potential,
which is the same as in \cite{SKS2007}.
This potential $a_1(r)$ is actually determined as the solution of the differential equation
\begin{equation}
 \frac{d^2 a_1}{dr^2} + \frac{d}{dr}\left(\Phi - \Lambda\right)\frac{da_1}{dr}
    - \frac{2}{r^2}e^{2\Lambda}a_1 = -4\pi e^{2\Lambda}j_1(r),
\end{equation}
where $j_1(r)$ is a radial component of the four-current given by
$j_1(r)=f_0r^2(\epsilon + p)$ where $f_0$ is a constant.
In order to determine the form of $a_1(r)$, we impose the regularity condition
at the stellar center, i.e., $a_1(r) = \alpha_c r^2 + O(r^4)$,
while at the stellar surface $a_1(r)$ should be smoothly connected to the exterior vacuum solution.

\section{Perturbation Equations}
\label{sec:III}

\subsection{Axisymmetric Polar Perturbations}

We restrict attention to axisymmetric polar-type perturbations with the relativistic Cowling
approximation, that is, the metric perturbations are ignored i.e. we set $\delta g_{\mu\nu}=0$.
Note that the polar type of perturbations are independent of the axial-type ones, which can be understood by recalling the  nature of the axisymmetric perturbations \citep{Lee2007}.
Then the Lagrangian displacement vector  for axisymmetric
polar-type perturbations can be written as
\begin{eqnarray}
 \xi^r      &=& \frac{e^{-\Lambda}}{r^2}W(t,r,\theta), \\
 \xi^\theta &=& -\frac{1}{r^2}V(t,r,\theta), \\
 \xi^\phi   &=& 0 \, .
\end{eqnarray}
Here we should note that the angular dependence of Lagrangian displacement vector $\xi^i$ for polar parity
can be in principle described via a spherical harmonic decomposition, i.e., $\xi^r\sim Y_{lm}$, $\xi^\theta\sim \partial_\theta Y_{lm}$, and
$\xi^\phi\sim \partial_\phi Y_{lm}/\sin^2\theta$. 
Then, the components of
the perturbed 4-velocity $\delta u^{\mu}$ in the Cowling approximation can be written as
\begin{eqnarray}
 \delta u^{t} &=& \delta u^{\phi} \  =\  0, \\
 \delta u^{r}      &=& \frac{1}{r^2}e^{-\Phi-\Lambda} \partial_t W, \\
 \delta u^{\theta} &=& -\frac{1}{r^2}e^{-\Phi} \partial_t V,
\end{eqnarray}
where $\partial_t$ denotes the partial derivative with respect to $t$.
Additionally, the Lagrangian variation of the baryon number density, $\Delta n$, is
defined as $\Delta n/n = -\nabla^{(3)}_k \xi^k$, where 
 $\nabla^{(3)}$ and $\Delta$ denote respectively the covariant derivative
in a 3-dimensional spatial part of the metric and the Lagrangian variation.
With the earlier definition for the Lagrangian displacement vector,
the Lagrangian variation of the baryon number density takes the form
\begin{equation}
 \frac{\Delta n}{n} = -\frac{e^{-\Lambda}}{r^2}\frac{\partial W}{\partial r}
       + \frac{1}{r^2}\left[\frac{\partial V}{\partial \theta} + \cot\theta V \right].
\end{equation}
Furthermore, for adiabatic perturbations, with the help of the first law of thermodynamics,
the perturbed density and pressure are given as
\begin{eqnarray}
 \delta \epsilon &=& (\epsilon + p)\frac{\Delta n}{n} - \frac{\partial \epsilon}{\partial r}\xi^r, \\
 \delta p        &=& \gamma p \frac{\Delta n}{n} - \frac{\partial p}{\partial r} \xi^r,
\end{eqnarray}
where $\gamma$ is the adiabatic constant defined as
\begin{equation}
 \gamma \equiv \left(\frac{\partial \ln p}{\partial \ln n}\right)_s
        = \left(\frac{\Delta p}{p}\right)\left( \frac{\Delta n}{n} \right)^{-1},
\end{equation}
and we also use the following relation between the Lagrangian perturbation
$\Delta$ and Eulerian perturbaiton $\delta$;
\begin{equation}
 \Delta f(t,r) \simeq \delta f(t,r) + \frac{\partial f}{\partial r}\xi^r.
\end{equation}
The above assumptions lead to the following form of the $r$- and $\theta$-components of the
linearized equation of motion with vanishing shear modulus, i.e., the equation (35) in \cite{SKS2007} with $\mu=0$,
\begin{eqnarray}
 \left(\epsilon + p + H^2\right) e^{-\Phi} \delta u^{r}_{\ ,t}
     &=& - \left(\delta \epsilon + \delta p\right)\Phi' e^{-2\Lambda} - e^{-2\Lambda}\delta p_{,r}
         + \left(\Lambda' + \frac{2}{r}\right)H^r\delta H^r + H^r \delta H^\alpha_{\ ,\alpha}
         + H^\theta \delta H^r_{\ ,\theta} + H^\phi \delta H^r_{\ ,\phi} \nonumber \\
     &&  + \left[-2\Phi'e^{-2\Lambda}H_\theta + H^r_{\ ,\theta} - 2re^{-2\Lambda}H^\theta
         + \cot\theta H^r - e^{-2\Lambda}H_{\theta,r}\right]\delta H^\theta
         - e^{-2\Lambda} H_\theta \delta H^\theta_{\ ,r} \nonumber \\
     &&  + \left[-2\Phi'e^{-2\Lambda}H_\phi + H^r_{\ ,\phi} -2re^{-2\Lambda}\sin^2\theta H^\phi
         - e^{-2\Lambda}H_{\phi,r}\right]\delta H^\phi - e^{-2\Lambda}H_\phi\delta H^\phi_{\ ,r}, 
     \label{perturbation-equation1} \\
 \left(\epsilon + p + H^2\right) e^{-\Phi} \delta u^{\theta}_{\ ,t}
     &=& - \frac{1}{r^2}\delta p_{,\theta} + \left[\left(\Phi' + \Lambda' + \frac{4}{r}\right)H^\theta
         + H^\theta_{\ ,r} - \frac{1}{r^2}H_{r,\theta}\right]\delta H^r
         + H^\theta \delta H^\alpha_{\ ,\alpha} - \frac{1}{r^2}H_r \delta H^r_{\ ,\theta} \nonumber \\
     &&  + \left[\left(\Phi' + \frac{2}{r}\right)H^r + \cot\theta H^\theta\right]\delta H^\theta
         + H^r \delta H^\theta_{\ ,r} + H^\phi \delta H^{\theta}_{\ ,\phi} \nonumber \\
     &&  + \left[\frac{1}{r^2}\left(H_{\theta,\phi} - H_{\phi,\theta}\right)
         - 2\sin\theta\cos\theta H^\phi\right]\delta H^\phi - \frac{1}{r^2}H_\phi \delta H^\phi_{\ ,\theta},
     \label{perturbation-equation2}
\end{eqnarray}
where a prime ($'$) denotes the derivative with respect to $r$, i.e., $'\equiv \partial_r$.
On the other hand, the linearized induction equation (37) in \cite{SKS2007} yields the
following relations for the first time-derivative of the components of $\delta H^\mu$
\begin{eqnarray}
 \delta H^t_{\ ,t} &=& e^{-\Phi}\left[H^{r}\delta u_{r,t} + H^{\theta}\delta u_{\theta,t}\right],
       \label{dHt} \\
 \delta H^r_{\ ,t} &=& -e^{\Phi}\left[\left\{\left(\Lambda' + \frac{2}{r}\right)H^r + H^r_{\ ,r}\right\}\delta u^{r}
     + \left(\cot\theta H^r + H^r_{\ ,\theta}\right)\delta u^{\theta}
     + H^r \delta u^{\theta}_{\ ,\theta} - H^\theta \delta u^{r}_{\ ,\theta}\right],
       \label{dHr} \\
 \delta H^{\theta}_{\ ,t} &=& -e^{\Phi}\left[\left\{\left(\Phi' + \Lambda' + \frac{2}{r}\right)H^{\theta}
     + H^\theta_{\ ,r} \right\} \delta u^{r} +  \left(\cot\theta H^{\theta} - \Phi' H^r
     + H^\theta_{\ ,\theta}\right)\delta u^{\theta}
     + H^\theta \delta u^{r}_{\ ,r} - H^{r} \delta u^{\theta}_{\ ,r}\right],
       \label{dHtheta} \\
 \delta H^{\phi}_{\ ,t} &=&  -e^{\Phi}\left[\left\{\left(\Phi' + \Lambda' + \frac{2}{r}\right)H^{\phi}
     + H^\phi_{\ ,r}\right\}\delta u^{r}
     + \left(\cot\theta H^{\phi} +  H^\phi_{\ ,\theta}\right)\delta u^{\theta}
     + H^{\phi} \left(\delta u^{r}_{\ ,r} + \delta u^{\theta}_{\ ,\theta}\right)\right].
       \label{dHphi}
\end{eqnarray}
Substituting the previous relations into the linearized equations
of motion (\ref{perturbation-equation1}) and (\ref{perturbation-equation2}),
one can obtain the following coupled system of evolution equations for the two functions describing the fluid perturbations
\begin{eqnarray}
 \left[
   \begin{array}{cc}
     {\cal A}_{00} & {\cal A}_{01} \\
     {\cal A}_{10} & {\cal A}_{11} \\
   \end{array}
 \right]
 \left[
   \begin{array}{c}
     \partial_t^2 W \\
     \partial_t^2 V \\
   \end{array}
 \right]
 =
 \left[
   \begin{array}{c}
     {\cal F}_W \\
     {\cal F}_V \\
   \end{array}
 \right],
 \label{ME}
\end{eqnarray}
where the coefficients ${\cal A}_{00}$, ${\cal A}_{01}$, ${\cal A}_{10}$, and ${\cal A}_{11}$
are 
\begin{eqnarray}
 {\cal A}_{00} &=& \left(\epsilon + p + H^\theta H_\theta + H^\phi H_\phi\right)
                   \frac{1}{r^2}e^{-2\Phi-\Lambda}, \\
 {\cal A}_{01} &=& \frac{1}{r^2} e^{-2\Phi}H^r H_\theta, \\
 {\cal A}_{10} &=& -\frac{1}{r^2} e^{-2\Phi - \Lambda} H^\theta H_r, \\
 {\cal A}_{11} &=& -\left(\epsilon + p + H^r H_r + H^\phi H_\phi\right) \frac{1}{r^2}e^{-2\Phi},
\end{eqnarray}
while the two expressions ${\cal F}_W$ and ${\cal F}_V$ are
\begin{eqnarray}
 {\cal F}_W 
     &=& - \left(\delta \epsilon + \delta p\right)\Phi' e^{-2\Lambda} - e^{-2\Lambda}\delta p_{,r}
         + \left(\Lambda' + \frac{2}{r}\right)H^r\delta H^r
         + H^r \left(\delta H^r_{\ ,r} + \delta H^\theta_{\ ,\theta} + \delta H^\phi_{\ ,\phi}\right)
         + H^\theta \delta H^r_{\ ,\theta} + H^\phi \delta H^r_{\ ,\phi} \nonumber \\
     &&  + \left[-2\Phi'e^{-2\Lambda}H_\theta + H^r_{\ ,\theta} - 2re^{-2\Lambda}H^\theta
         + \cot\theta H^r - e^{-2\Lambda}H_{\theta,r}\right]\delta H^\theta
         - e^{-2\Lambda} H_\theta \delta H^\theta_{\ ,r} \nonumber \\
     &&  + \left[-2\Phi'e^{-2\Lambda}H_\phi + H^r_{\ ,\phi} -2re^{-2\Lambda}\sin^2\theta H^\phi
         - e^{-2\Lambda}H_{\phi,r}\right]\delta H^\phi - e^{-2\Lambda}H_\phi\delta H^\phi_{\ ,r}, \\
 {\cal F}_V 
     &=& - \frac{1}{r^2}\delta p_{,\theta} + \left[\left(\Phi' + \Lambda' + \frac{4}{r}\right)H^\theta
         + H^\theta_{\ ,r} - \frac{1}{r^2}H_{r,\theta}\right]\delta H^r
         + H^\theta \left(\delta H^r_{\ ,r} + \delta H^\theta_{\ ,\theta} + \delta H^\phi_{\ ,\phi}\right)
         - \frac{1}{r^2}H_r \delta H^r_{\ ,\theta} \nonumber \\
     &&  + \left[\left(\Phi' + \frac{2}{r}\right)H^r + \cot\theta H^\theta\right]\delta H^\theta
         + H^r \delta H^\theta_{\ ,r} + H^\phi \delta H^\theta_{\ ,\phi} \nonumber \\
     &&  + \left[\frac{1}{r^2}\left(H_{\theta,\phi} - H_{\phi,\theta}\right)
         - 2\sin\theta\cos\theta H^\phi\right]\delta H^\phi - \frac{1}{r^2}H_\phi \delta H^\phi_{\ ,\theta}.
\end{eqnarray}
Notice that  the above system of evolution equations has been written 
for a general equilibrium magnetic field $H^\mu(r,\theta,\phi)$, i.e.,
the form of the magnetic field has not yet been specified.

\subsection{Dipole magnetic field}

As we mentioned earlier, in this article  we  study the perturbations of  dipole fields.
In order to simplify the boundary condition at the stellar surface,
we introduce two new functions $w(t,r,\theta)$ and $v(t,r,\theta)$
defined as 
\begin{eqnarray}
 w &=& \epsilon W \sin\theta,  \label{eq:trans1} \\
 v &=& \epsilon V \sin\theta \, . \label{eq:trans2}
\end{eqnarray}
Notice that in the new functions the trigonometric term,  $\sin\theta$, has been introduced in an attempt to 
improve the numerical stability of the code. 
Then, by restricting our attention only  to axisymmetric dipole poloidal magnetic fields describe by (\ref{DMF}),
we get a simplified form of the linearized equation of motion (\ref{ME}), that is
\begin{eqnarray}
 \left[
   \begin{array}{cc}
     {\cal A}_{00} & {\cal A}_{01} \\
     {\cal A}_{10} & {\cal A}_{11} \\
   \end{array}
 \right]
 \left[
   \begin{array}{c}
     \partial_t^2 w \\
     \partial_t^2 v \\
   \end{array}
 \right]
 =
 \epsilon\sin\theta
 \left[
   \begin{array}{c}
     {\cal F}_W \\
     {\cal F}_V \\
   \end{array}
 \right], \label{evolution}
\end{eqnarray}
where the coefficients ${\cal A}_{00}$, ${\cal A}_{01}$, ${\cal A}_{10}$, and ${\cal A}_{11}$
become
\begin{eqnarray}
 {\cal A}_{00} &=& e^{-2\Phi - \Lambda}
                   \left[\epsilon + p + \frac{1}{4\pi r^2} \left(e^{-\Lambda}{a_1}' \sin\theta\right)^2
                   \right], \\
 {\cal A}_{01} &=& -\frac{a_1 {a_1}'}{2\pi r^2}e^{-2\Phi - 2\Lambda} \sin\theta\cos\theta, \\
 {\cal A}_{10} &=& \frac{a_1 {a_1}'}{2\pi r^2}e^{-2\Phi - \Lambda} \sin\theta\cos\theta, \\
 {\cal A}_{11} &=& -e^{-2\Phi} \left[(\epsilon + p)r^2 + \frac{{a_1}^2}{\pi r^2}\cos^2\theta\right],
\end{eqnarray}
and the expressions ${\cal F}_W$ and ${\cal F}_V$ will be simplified to the following forms
\begin{eqnarray}
 {\cal F}_W 
     &=& - \left(\delta \epsilon + \delta p\right)\Phi' e^{-2\Lambda} - e^{-2\Lambda}\delta p_{,r}
         + \frac{e^{-\Lambda}}{\sqrt{4\pi}}\left[-4\pi j_1\delta H^\theta
         + {a_1}'\left\{\left(\Phi' + \frac{2}{r}\right)e^{-2\Lambda}
           \delta H^\theta + e^{-2\Lambda}\delta H^\theta_{\ ,r} - \frac{1}{r^2}\delta H^r_{\ ,\theta}
           \right\}\right]\sin\theta, \\
 {\cal F}_V 
     &=& - \delta p_{,\theta} + \sqrt{4\pi}e^{\Lambda}j_1
           \delta H^r \sin\theta 
         + \frac{e^{-\Lambda}a_1}{\sqrt{\pi}}\left[\left(\Phi' + \frac{2}{r}\right)\delta H^\theta
         + \delta H^\theta_{\ ,r} - \frac{1}{r^2}e^{2\Lambda}\delta H^r_{\ ,\theta}\right]\cos\theta.
\end{eqnarray}
Notice that in order to derive the above equations, we used the relation
\begin{equation}
 \left(\Lambda' + \frac{2}{r}\right)\delta H^r 
         + \delta H^r_{\ ,r} + \delta H^\theta_{\ ,\theta}
         + \cot\theta \delta H^\theta = 0.
\end{equation}
This relation is satisfied throughout the star for the axisymmetric polar perturbations.

The components of
the perturbed magnetic field $\delta H^k$ defined in equations (\ref{dHr}) -- (\ref{dHphi}) can be written in terms of the  new perturbation functions described by (\ref{eq:trans1}) and (\ref{eq:trans2}), as
\begin{eqnarray}
 \delta H^r
    &=& - \frac{e^{-\Lambda}}{\sqrt{4\pi}r^2\epsilon\sin\theta}\left[e^{-\Lambda}{a_1}'
          \left(w\cos\theta + w_{,\theta}\sin\theta \right)
        + 2a_1 \left(v\sin\theta - v_{,\theta} \cos\theta\right)\right], \\
 \delta H^\theta 
    &=& - \frac{e^{-\Lambda}}{\sqrt{4\pi}r^2\epsilon\sin\theta} \bigg[\left\{\left(\Phi' + \frac{2}{r}
        + \frac{\epsilon'}{\epsilon}\right){a_1}' - \frac{2a_1}{r^2} e^{2\Lambda}
        + 4\pi e^{2\Lambda}j_1\right\}e^{-\Lambda} w\sin\theta \nonumber \\
    &&  \hspace{2.5cm} + 2 v\left({a_1}' - \frac{2a_1}{r} - \frac{a_1 \epsilon'}{\epsilon}\right)\cos\theta
        - e^{-\Lambda}{a_1}' w'\sin\theta + 2a_1 v'\cos\theta\bigg],
\end{eqnarray}
with $\delta H^\phi=0$. 

Finally, the variations of pressure and density are described by
\begin{eqnarray}
 \delta p &=& \frac{\gamma p}{\epsilon\sin\theta}\left(-e^{-\Lambda}w' + v_{,\theta}\right)
           + \frac{e^{-\Lambda}}{\epsilon\sin\theta}\left(\frac{\gamma p\epsilon'}{\epsilon}-p'\right)w, \\
 \delta \epsilon
          &=& \frac{\epsilon + p}{\epsilon\sin\theta}\left(-e^{-\Lambda}w' + v_{,\theta}\right)
           + \frac{p\epsilon'}{\epsilon^2\sin\theta}e^{-\Lambda}w.
\end{eqnarray}

\subsection{Boundary Conditions}

For axisymmetric polar perturbations, in general,
the angular dependences of the functions $w(t,r,\theta)$ and $v(t,r,\theta)$ can be written
as $w\propto \sin\theta P_\ell(\cos\theta)$ and
$v\propto \sin\theta\, \partial_\theta P_\ell(\cos\theta)$, where
$P_\ell(\cos\theta)$ is the Legendre polynomial of order $\ell$.
Thus the boundary conditions on the axis ($\theta=0$) will be
\begin{equation}
 w=0\ \ \ {\mbox{and}}\ \ \ v=0,
\end{equation}
while on the equatorial plane ($\theta=\pi/2$) we set
\begin{eqnarray}
 w=0\ \ \ {\mbox{and}}\ \ \ v_{,\theta}=0\ \ \ &&{\mbox{for odd $\ell$}},  \label{eq:odd-BC} \\
 w_{,\theta}=0\ \ \ {\mbox{and}}\ \ \ v=0\ \ \ &&{\mbox{for even $\ell$}}. \label{eq:even-BC}
\end{eqnarray}
Finally, at the stellar center we impose the regularity conditions for $w$ and $v$,
i.e., $w=0$ and $v=0$ at $r=0$. While the boundary condition at the stellar surface is
that the Lagrangian perturbation of the pressure should be zero, i.e., $\Delta P=0$,
which is equivalent to $w=0$ and $v=0$ as we will use a barotropic equation of state and
$\epsilon=0$ at the stellar surface for the background.

\section{Numerical Results}
\label{sec:IV}

\subsection{Code Tests}

For our studies 
we adopted a polytropic equation of state (EoS) defined as
\begin{equation}
  p = Kn_0m_b\left(\frac{n}{n_0}\right)^\Gamma\ \ \ \ {\rm and}
  \ \ \ \ \ \epsilon = nm_b + \frac{p}{\Gamma-1},
\end{equation}
where $m_b$ and $n_0$ are the baryon mass and normalized number density given by
$m_b=1.66\times 10^{-24}$ g and $n_0=0.1$ fm$^{-3}$, respectively.
Specifically, we adopt the values of $\Gamma$ and $K$ as $\Gamma=2.46$ and $K=0.00936$,
whose values are in good agreement with the tabulated data for the realistic EoS A \citep{EoS_A}.
For this EoS we focus only on the stellar model with mass $M=1.4M_\odot$, radius $R=10.35$ km,
and compactness $M/R=0.2$.

For the time evolution of the perturbation equations (\ref{evolution}),
we use the iterated Crank-Nicholson scheme \citep{CN}, which keeps second-order
accuracy for the both of space and time. In order to eliminate the spurious higher frequencies,
we also add a 4th-order Kreiss-Oliger dissipation into the evolution equations \citep{GKO1995},
where the coefficient of this dissipation term, $\varepsilon_D$, depends on the numerical grid size. 
This type of numerical viscosity has been used successfully in \cite{SKS2008} and \cite{Gaertig2008}. 
Most of the numerical calculations in this paper were done in the 2D region of $(r,\theta)$ with $50\times40$
grid points, where $\theta$ is in the range of $0\le\theta\le\pi/2$.
The values of $\theta =0$ and $\pi/2$ correspond to the axis and the equatorial plane, respectively.
It should be mentioned  that the results were not significantly improved when a larger number of grid points has been used, i.e., $50\times80$ or $100\times40$.
The numerical evolutions were stable for long enough to allow us to   extract the specific frequencies for each spatial point by performing
a Fast Fourier Transformation (FFT) of the time varying perturbation functions. 
Since in this paper we evolve the perturbed models for about $T=200$ msec, in the FFT
we can expect that possible maximum error in the estimation of the frequency $\Delta f$,
is around $\Delta f\sim 1/T \simeq 5$ Hz.

Before studying in detail  the magnetar oscillations we performed test of the 2D evolution code 
on  non-magnetized stellar models.
In general for a spherically symmetric star which is not endowed with a magnetic field,
the oscillations can be decomposed into spherical harmonics $Y_{\ell m}$
which are independent of the angular index $\ell$. Thus by using the appropriate boundary conditions
we can determine the specific frequencies by transforming it into an eigenvalue problem.
The idea here is to compare the frequencies determined by the eigenvalue methods, to those derived from our evolution code in order to check the accuracy of the new method. For the solution of the eigenvalue problem
we use the same code as in \cite{Sotani2004}.
The regularity condition at stellar center, demands that the behavior of $W$ and $V$ in the
vicinity of stellar center is of the form
$W\sim Cr^{\ell + 1}+ \cdots$ and $V\sim -C r^\ell/\ell + \cdots$, where $C$ is some constant.
Thus we set the initial data for the time evolution in the form
\begin{eqnarray}
 w(r,\theta) &=& \epsilon r^{l+1} \sin\theta P_\ell(\cos\theta), \label{eq:wi} \\
 v(r,\theta) &=& -\frac{\epsilon}{\ell} r^\ell \sin\theta\, \partial_\theta P_{\ell}(\cos\theta).
                 \label{eq:vi}
\end{eqnarray}
On a non-rotating cold neutron star without magnetic fields
there exists only one family of fluid mode, i.e., the $p$ (pressure) modes.
The $f$ mode is a fundamental mode of the pressure mode family, i.e., the $p_0$ mode.
There exists only one such mode, while there is an infinite number of higher $p$ modes for every $\ell$.
Traditionally the $p$ modes are named as $p_1$, $p_2$, $\dots$ in ascending order of those frequencies.
Note that the subscript used in the notation of the $p$ modes shows in addition the number of
radial nodes in the eigenfunction, while the eigenfunction of the $f$ mode has no nodes.
In table \ref{Tab:test}, we can compare the frequencies determined by the solution of the eigenvalue problem and
those derived by evolving the 2D perturbation equations. We list results only for the $f$, $p_1$ and $p_2$ modes
with harmonic index $\ell=2$, $3$ and 4.
Since the oscillations with lower frequencies (or with longer wave-length)
are not so sensitive to the number of grid points, the lower frequencies is expected to be calculated with higher accuracy.
Actually, we find that the $f$ mode with $\ell=2$ shows a deviation from the values derived by the eigenvalue problem of the order of $2.6$\%
while the $f$ mode with $\ell=4$ shows a deviation of about $3.7$\%.
In any case the results derived with the time evolution code show a maximum deviation which is always bellow 5\%
even for the higher frequency modes. We consider that this accuracy is acceptable for the type of study that
we perform here, i.e., to quantify the effect of the magnetic field on the fluid modes and to find out
whether the polar Alfv\'{e}n oscillations are discrete or continuous.

\begin{table*}
 \centering
  \caption{The frequencies for $f$, $p_1$ and $p_2$ modes, for the stellar model with $M=1.4M_\odot$,
  for the harmonic indices $\ell=2$, $3$ and $4$
  in kHz. Here the superscripts (E) and (T) denote the frequencies found by solving the
  eigenvalue problem and those by evolving the 2-dimensional perturbation equations, respectively.}
\label{Tab:test}
  \begin{tabular}{ccccccc}
  \hline
   $\ell$ & $f^{\rm (E)}$ & $f^{\rm (T)}$ & $p_{1}^{\rm (E)}$ & $p_{1}^{\rm (T)}$ &
   $p_{2}^{\rm (E)}$ & $p_{2}^{\rm (T)}$ \\
 \hline
 2 & 2.68 & 2.75 & 6.71 & 6.87 &  9.97 & 10.21 \\
 3 & 3.26 & 3.41 & 7.65 & 7.89 & 11.07 & 11.38 \\
 4 & 3.75 & 3.89 & 8.49 & 8.73 & 12.08 & 12.38 \\
\hline
\end{tabular}
\end{table*}
%
%
%

\subsection{Polar Oscillations of Magnetars}

As initial data for the evolution of the perturbation equations we used the analytic forms for the perturbation functions  $w(r,\theta)$ and $v(r,\theta)$ described by the equations (\ref{eq:wi}) and (\ref{eq:vi}). We set initial data for angular indices $\ell=2$, $3$ and $4$ and we 
imposed at the equatorial plane the boundary conditions described by the equations (\ref{eq:odd-BC}) and (\ref{eq:even-BC}).
Actually, the perturbations on an axisymmetric background can be decomposed into two classes
that satisfy different conditions at the equatorial plane, i.e., one class satisfies the condition
specified in equation (\ref{eq:odd-BC}) and the other one satisfies the condition given by
equation (\ref{eq:even-BC}). These conditions describe perturbations for even and odd values of the angular index $\ell$.

\begin{figure}
\begin{center}
\begin{tabular}{ccc}
\includegraphics[scale=0.33]{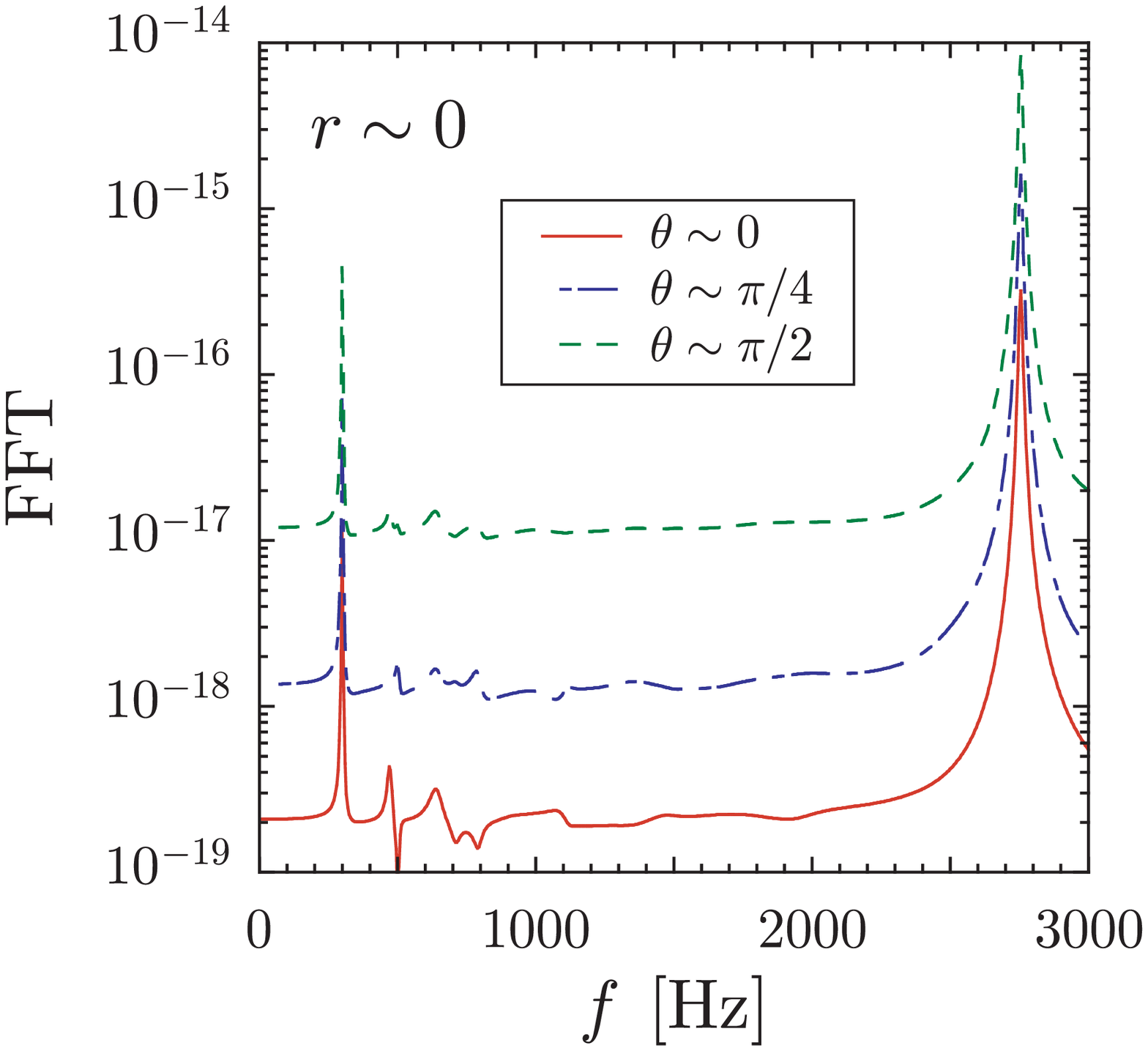} &
\includegraphics[scale=0.33]{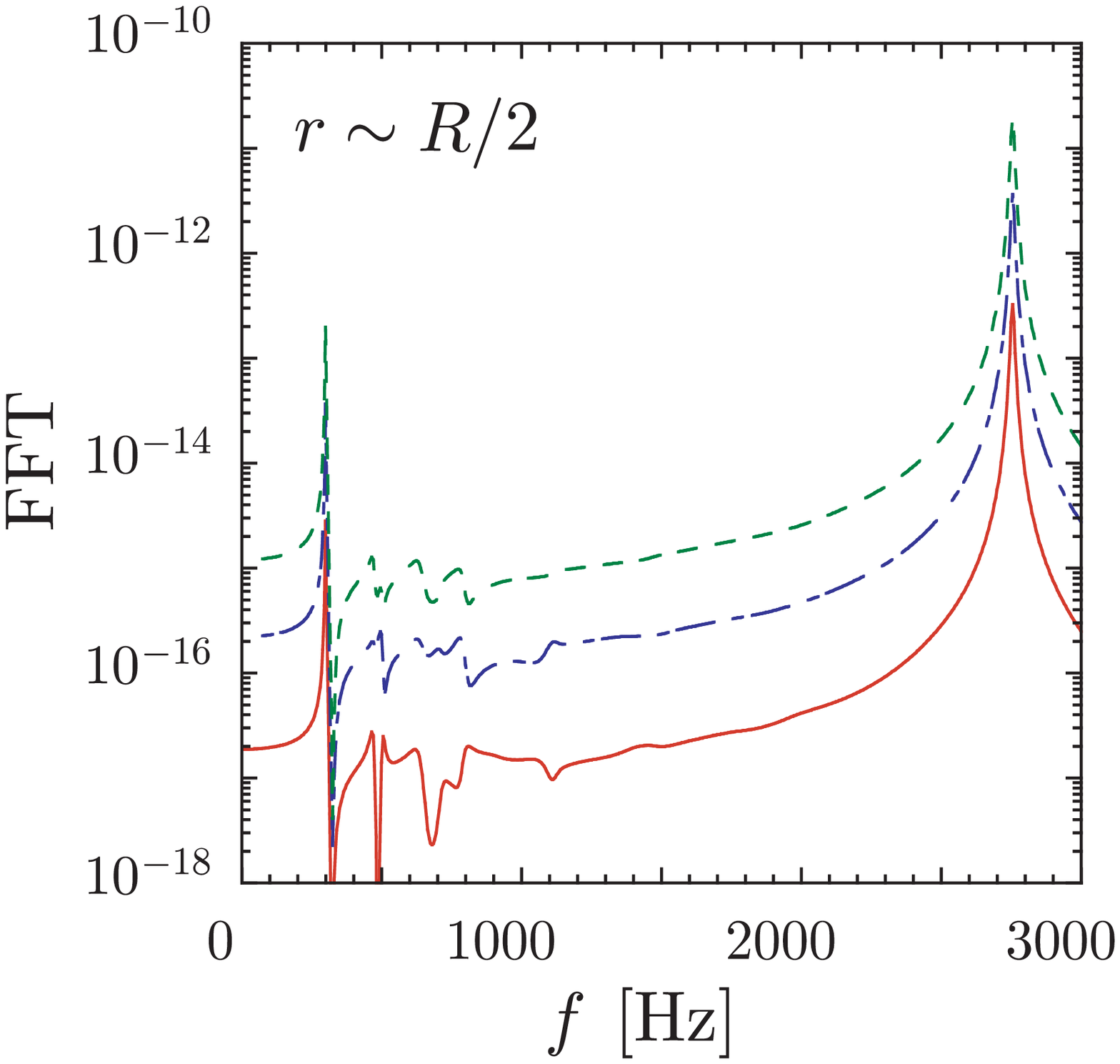} &
\includegraphics[scale=0.33]{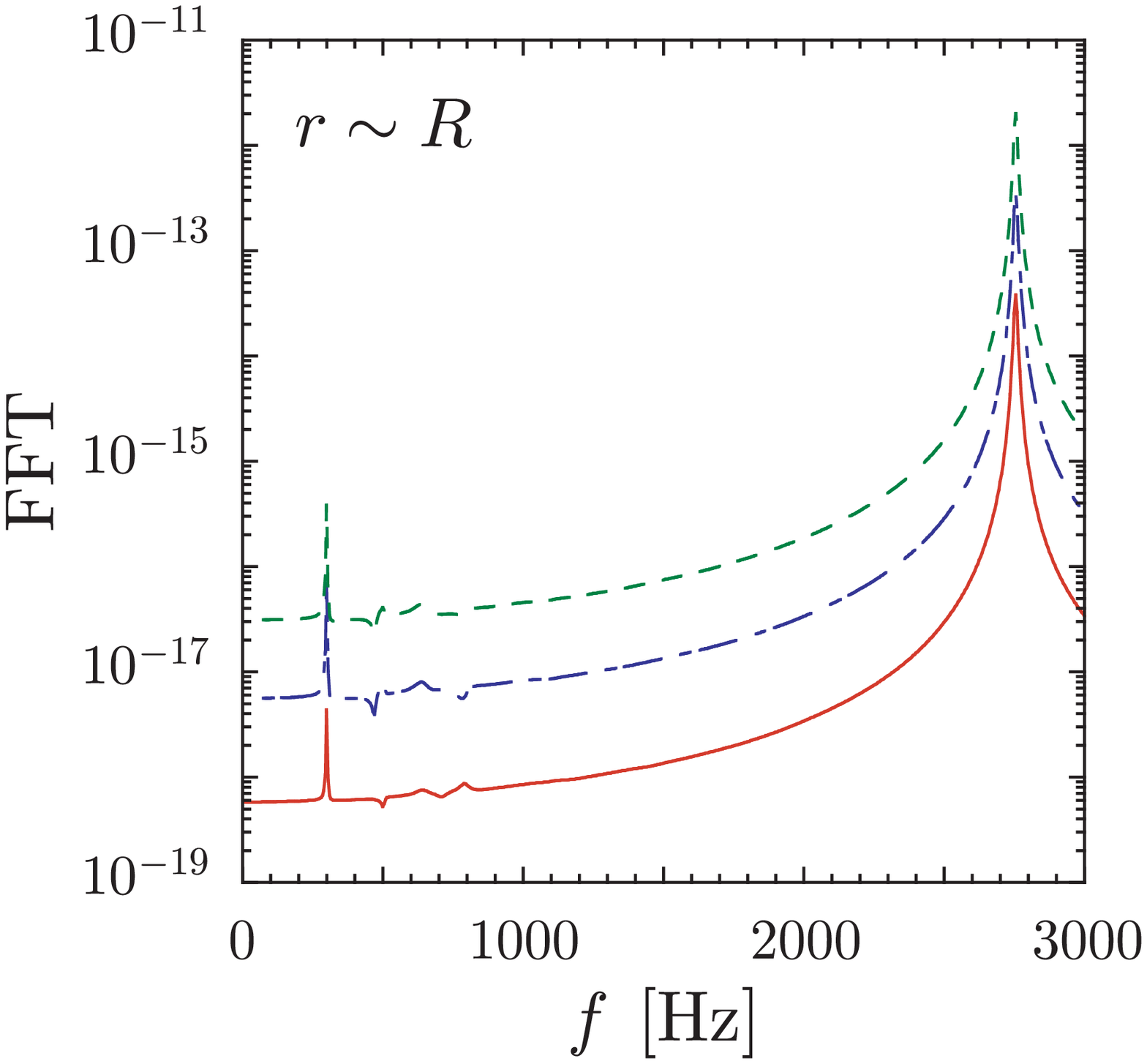} \\
\end{tabular}
\caption{
The FFT of the perturbation function $v(t,r,\theta)$ for a stellar model with $B=1\times 10^{16}$ Gauss.
The three lines in each figure  correspond
to three different angular positions, $\theta\sim0$, $\pi/4$, and $\pi/2$, and the different
figures correspond to the different radial positions, $r\sim 0$, $R/2$, and $R$.
In all figures we can see one peak for the Alfv\'{e}n oscillations,
which is corresponding to ${}_2a_0$ mode (i.e. the fundamental $l=2$ polar Alfv\'{e}n mode), and another
peak at higher frequencies corresponding to the frequency of the $f$ mode with $\ell=2$ (${}_2f$).
}
\label{fig:FFT}
\end{center}
\end{figure}
%
%
%
%
\begin{figure}
\begin{center}
\begin{tabular}{cc}
\includegraphics[scale=0.42]{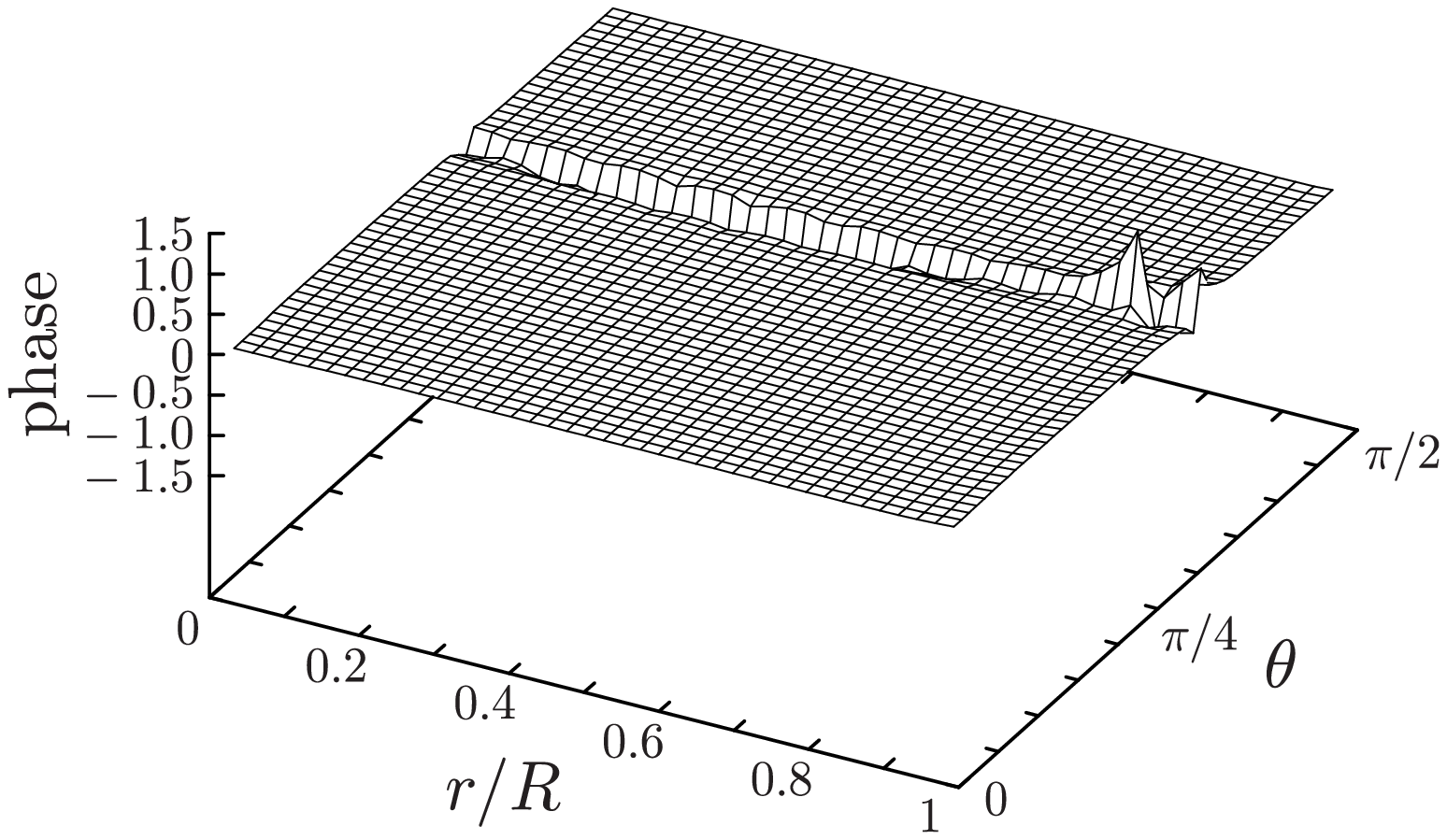} &
\includegraphics[scale=0.42]{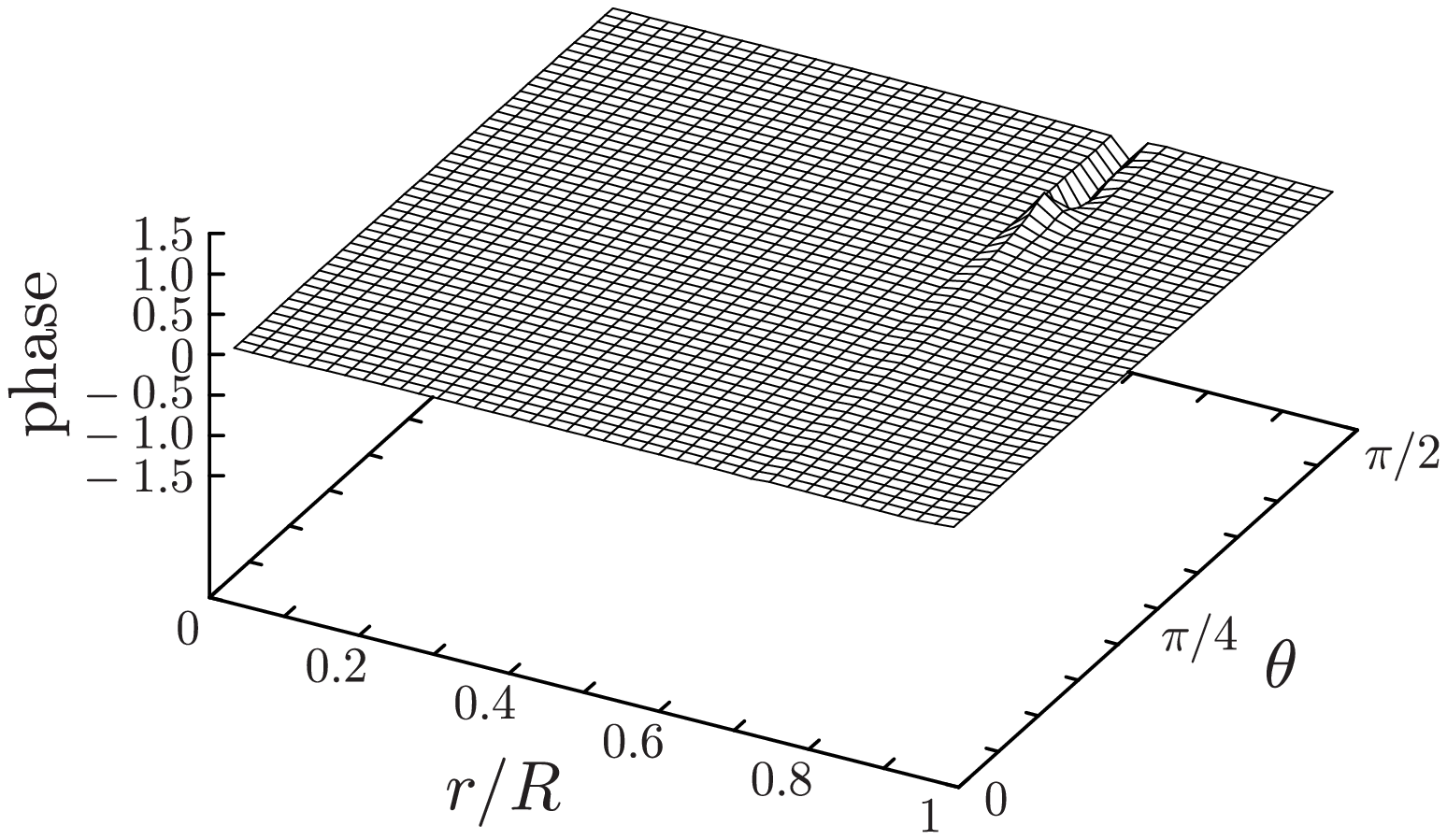} \\
\end{tabular}
\caption{
The phase of the fundamental polar Alfv\'{e}n mode with $\ell=2$ (${}_2a_0$) for a magnetar with $B=1\times 10^{16}$ Gauss.
The left and right panels correspond to the perturbation functions $w(t,r,\theta)$ and $v(t,r,\theta)$,
respectively.
}
\label{fig:phase}
\end{center}
\end{figure}

\begin{figure}
\begin{center}
\begin{tabular}{ccc}
\includegraphics[scale=0.42]{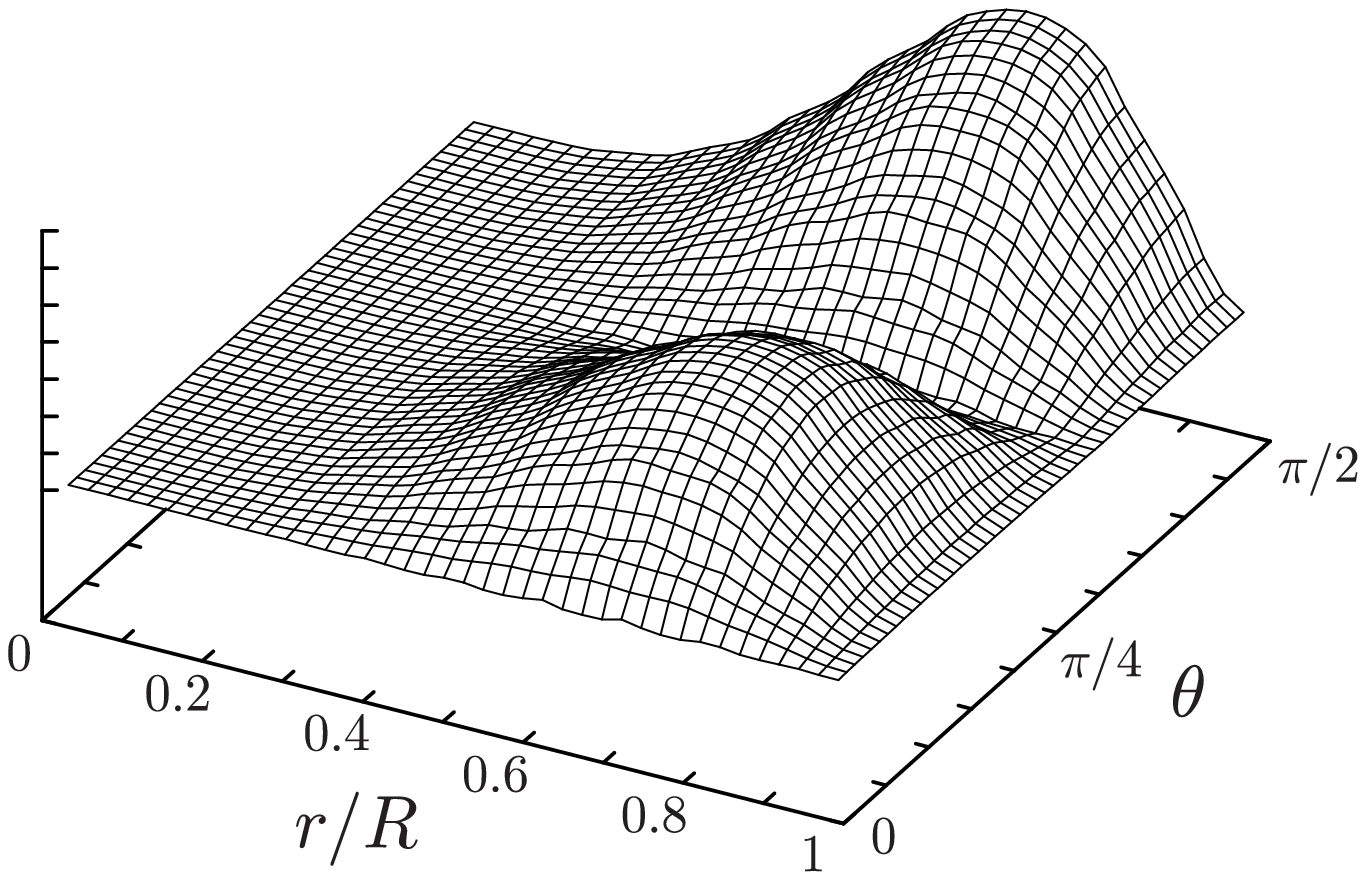} &
\includegraphics[scale=0.42]{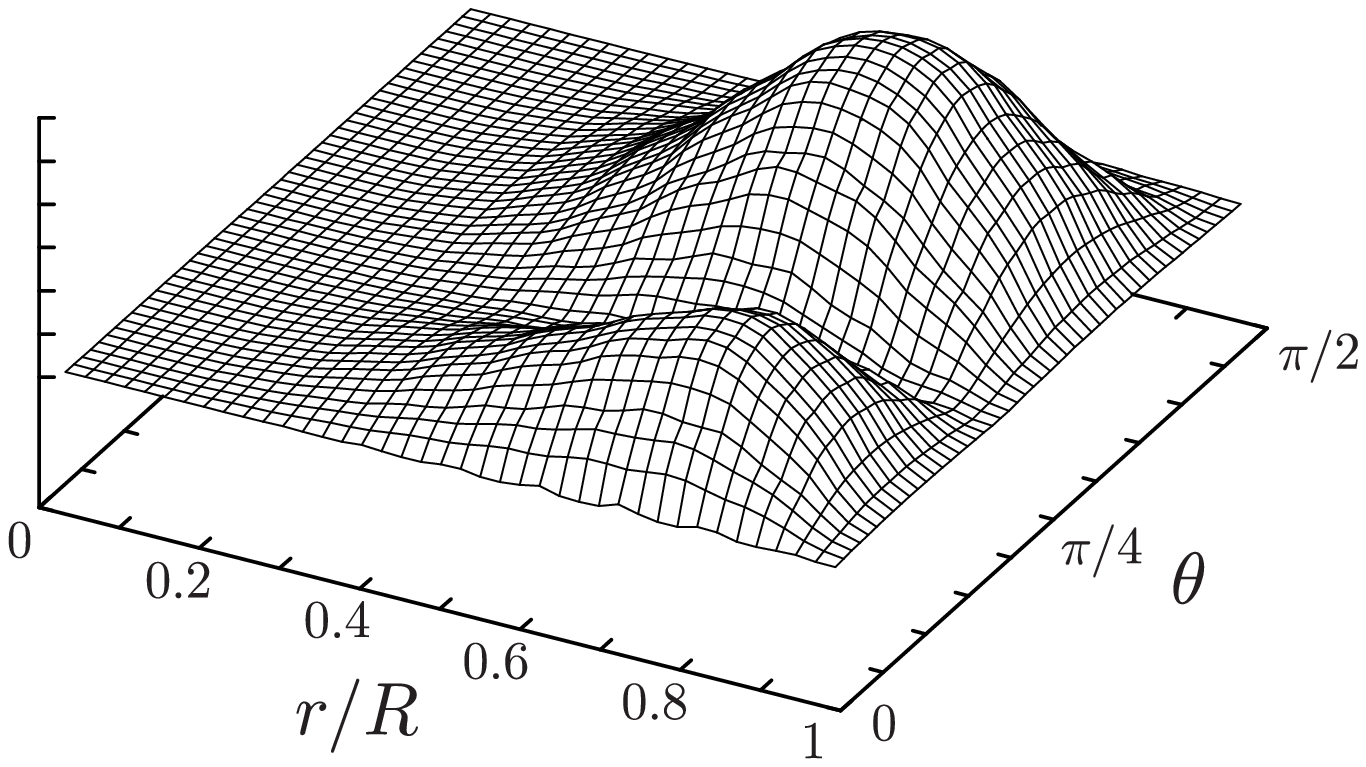} &
\includegraphics[scale=0.42]{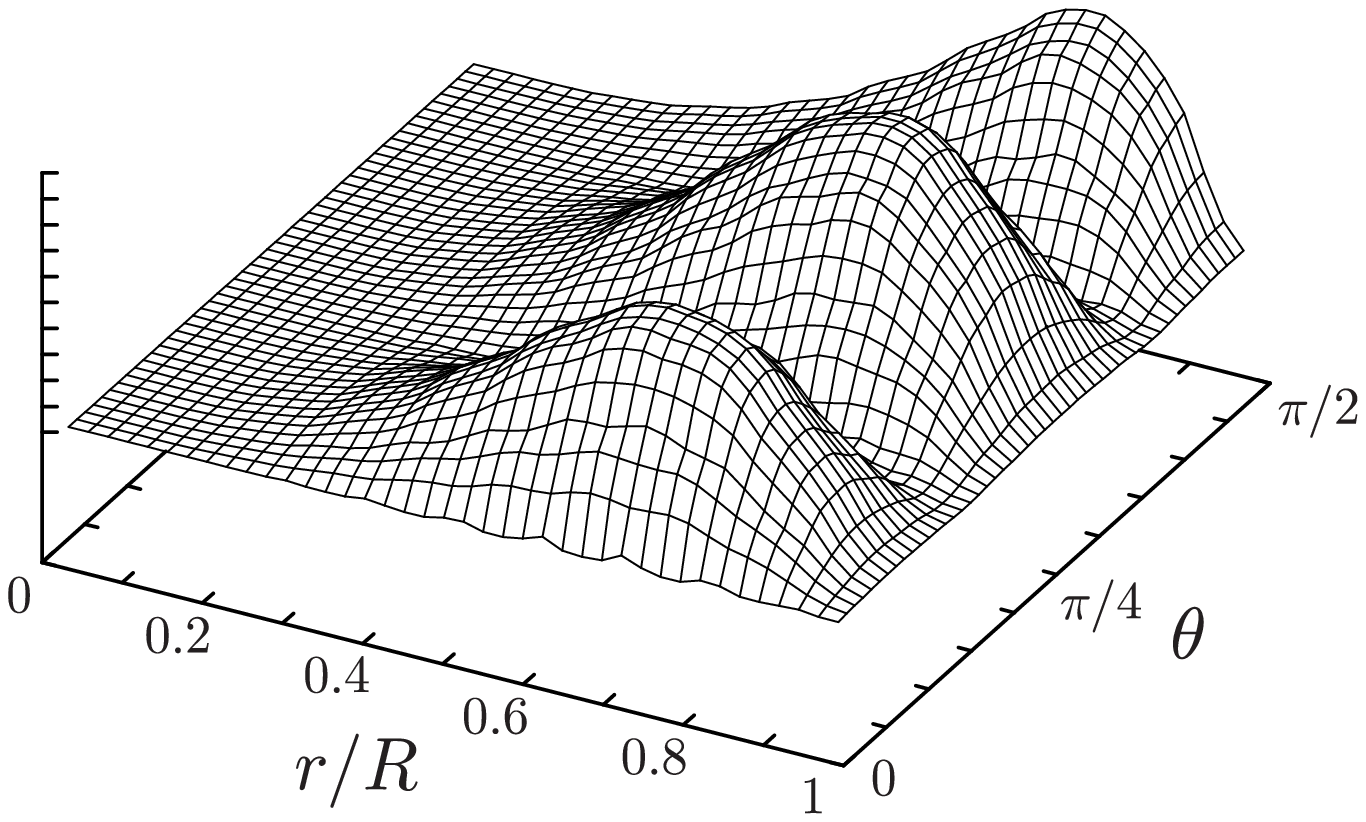} \\
\end{tabular}
\caption{
Effective amplitude of the eigenfunction for $w(t,r,\theta)$ for polar Alfv\'{e}n
oscillation modes. The three panels correspond to  Alfv\'{e}n oscillations
with $\ell=2$ (left panel), with $\ell=3$ (center panel), and with $\ell=4$ (right panel).
}
\label{fig:eigenfunction-w}
\end{center}
\end{figure}

\begin{figure}
\begin{center}
\begin{tabular}{ccc}
\includegraphics[scale=0.42]{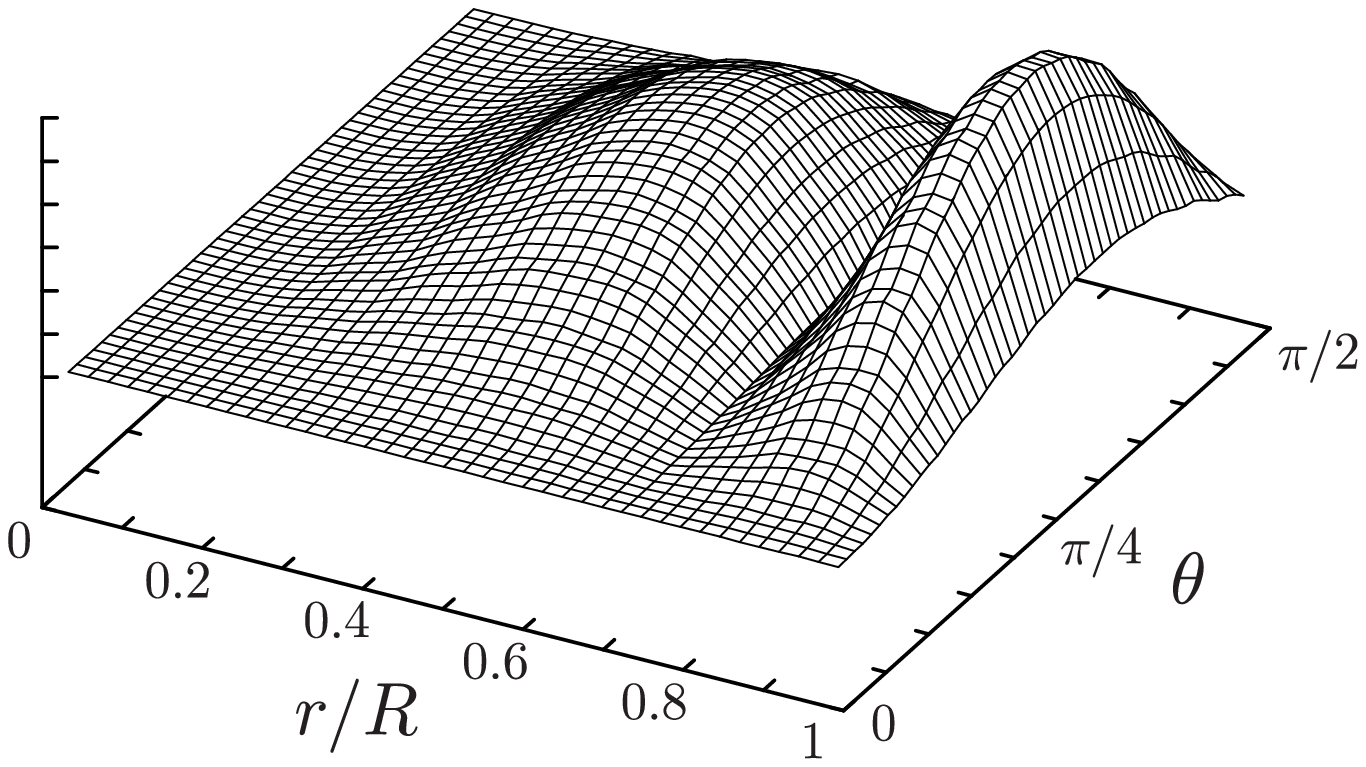} &
\includegraphics[scale=0.42]{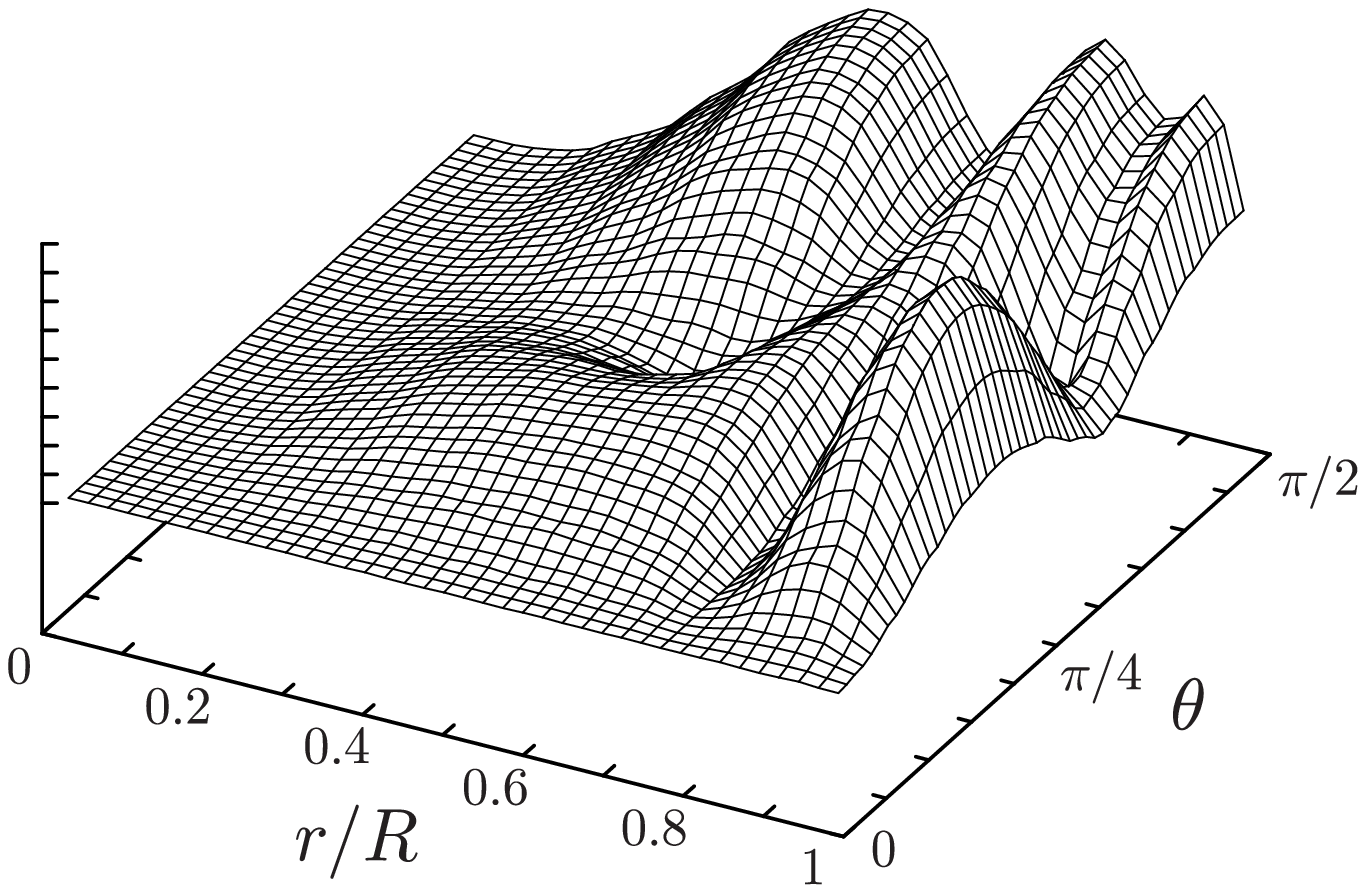} &
\includegraphics[scale=0.42]{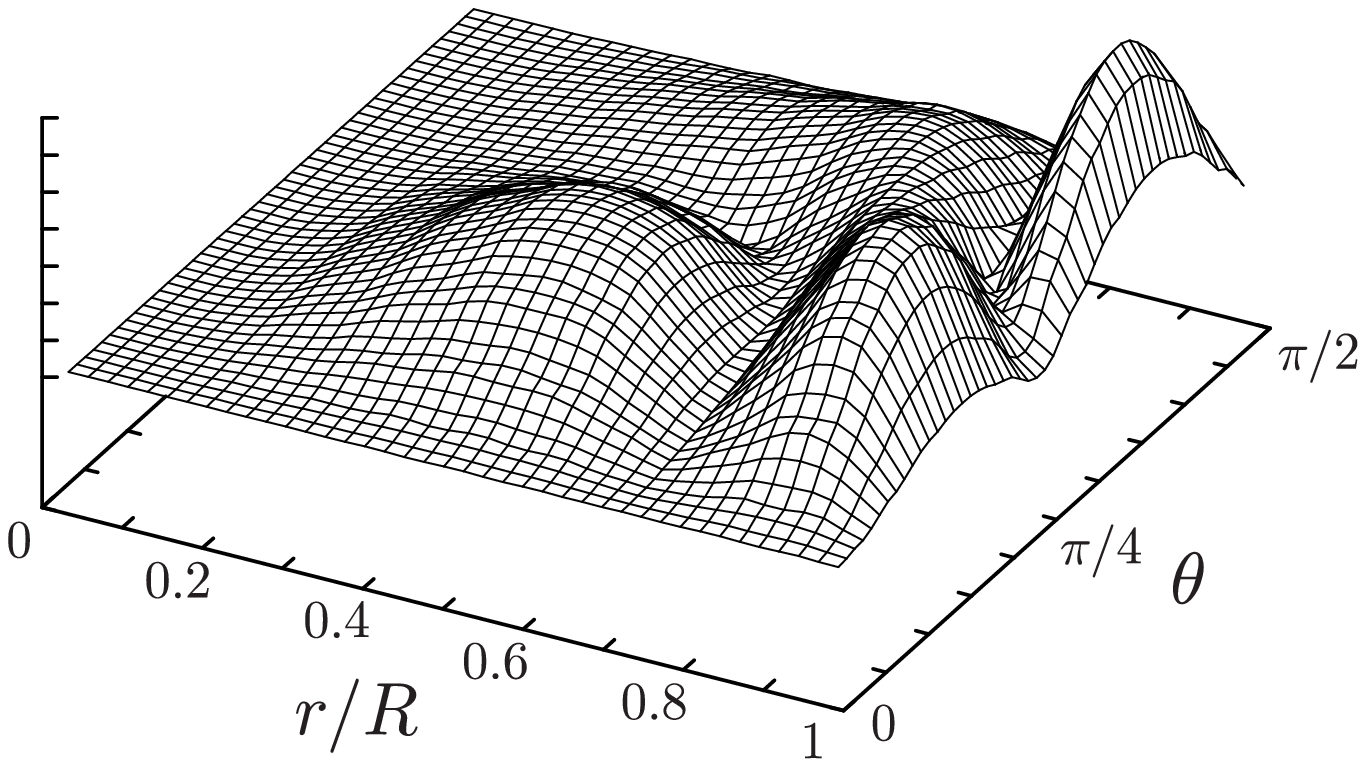} \\
\end{tabular}
\caption{
Effective amplitude of the eigenfunction for $v(t,r,\theta)$ with the several Alfv\'{e}n
oscillation modes. The three panels are corresponding to the Alfv\'{e}n oscillations
with $\ell=2$ (left panel), with $\ell=3$ (center panel), and with $\ell=4$ (right panel).
}
\label{fig:eigenfunction-v}
\end{center}
\end{figure}

The various features of those oscillations are examined in two ways.
The first involves checking of the FFT amplitude at various points inside the star.
If the spectrum is continuous the peaks in FFT will depend on the position,
while if the peaks are independent of the observer's position then they correspond to a
discrete spectrum. The second check involves the study of the phase of oscillation
for each peak frequency in FFT, i.e., for the discrete spectrum the phase should be constant throughout the star.
Figure \ref{fig:FFT} shows
the FFT of the perturbation function   $v(t,r,\theta)$ at various points inside the star
with $B=1\times 10^{16}$ Gauss. The time-series has been generated by evolving an initial data set corresponding to $\ell=2$ deformations.
In this figure the three different
lines  correspond to different angular positions i.e., $\theta\sim 0$, $\pi/4$, and $\pi/2$,
while the different panels correspond to different radial positions i.e., $r\sim 0$, $R/2$, and $R$.
It should be noted that the FFT of the other perturbation function, $w(t,r,\theta)$ shows
 that the peaks correspond exactly to the same frequencies as in figure \ref{fig:FFT} and obviously the peaks are
independent of the observer's position. 
The peak at 2754 Hz corresponds to the ${}_2f$ mode,
while the lower peak at 300 Hz corresponds to the polar Alfv\'{e}n oscillation mode ${}_2a_0$.
Next we study the phase of the specific Alfv\'{e}n oscillation mode ${}_2a_0$.
In figure \ref{fig:phase} we show the phase of the oscillation mode ${}_2a_0$
for a magnetar with $B=1\times 10^{16}$ Gauss. The left and right panels  correspond
to the phases for the perturbation functions $w(t,r,\theta)$ and  $v(t,r,\theta)$, respectively.
We observe that the phases for both functions $w$ and $v$ are almost constant except for some regions
around $\theta\sim \pi/3$ for the function $w$ and a part of $r/R\sim 0.8$ for the function $v$.
We identify that these two regions with somehow strange behavior of the phase are related to the due to numerical loss of accuracy in the calculation of the effective amplitudes,
where the eigenfunctions are almost zero because these are nodal points (see the left panels in figures \ref{fig:eigenfunction-w}
and \ref{fig:eigenfunction-v}).
Thus we can argue that the phase of the oscillation modes is independent from the position.
With the above two tests we conclude that the Alfv\'{e}n oscillations of polar
parity are described by a discrete spectrum, in contrast to the spectrum of the axial Alfv\'{e}n oscillations.
This feature seems to be very similar to the case of inertial modes
of rotating stars.
That is, the spectrum of the axial inertial modes seems to be continuous, at least in the slow rotation approximation (e.g., \cite{Kojima1998,BK1999}), while
the polar  modes admit a discrete spectrum.

%
%
%
%

Different initial data sets, such as initial deformation for $\ell=3$ and $4$,
verify the previous results, i.e., the spectrum is discrete and the oscillation modes have  constant phase.
In figures \ref{fig:eigenfunction-w} and \ref{fig:eigenfunction-v}, 
we show the effective amplitudes of the eigenfunctions  $w(r,\theta)$ and $v(r,\theta)$
with the corresponding peak frequencies in FFT,
 300 Hz (for $\ell=2$), 465 Hz (for $\ell=3$), and 500 Hz (for $\ell=4$),
where we assumed that magnetic field strength is $B=1\times 10^{16}$ Gauss.

\begin{figure}
\begin{center}
\includegraphics[height=6cm]{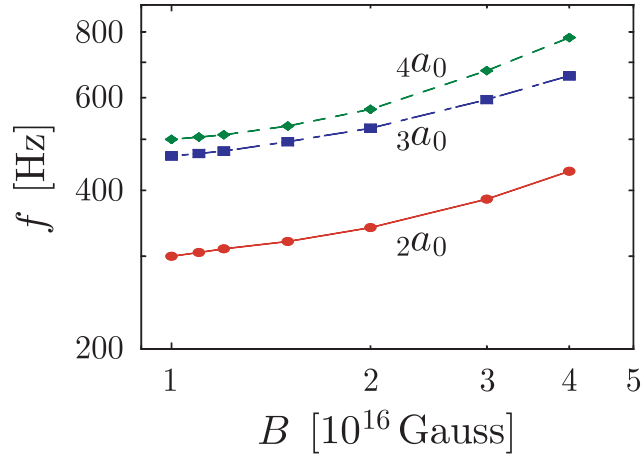}
\caption{
The frequencies of the Alfv\'{e}n modes as functions of magnetic field strength.}
\label{fig:Alfven}
\end{center}
\end{figure}

In figure \ref{fig:Alfven}, we show the results of our examination for the  dependence of the polar Alfv\'{e}n modes
on magnetic field strength.
This dependence of the polar Alfv\'{e}n oscillations on the magnetic field strength, can be eventually used
to explain the observed higher QPO frequencies such as 150, 625 and 1840 in SGR 1806--20
or 155 Hz in SGR 1900+14.
It should be noted that for weaker magnetic fields 
it is difficult to determine the frequencies of the Alfv\'{e}n
modes because the corresponding peaks in FFT
become a few orders of magnitude smaller compared to those of the fluid modes.
This observation suggests that the polar Alfv\'{e}n oscillations will play no role on the dynamics of magnetars with weaker magnetic fields.
In order to find the dependence  of the polar Alfv\'{e}n mode frequencies
to moderate and weak magnetic field strengths, one may study the problem by using mode analysis 
i.e. by decomposing the perturbations into spherical harmonics $Y_{\ell m}$. This type of study has already been done for Newtonian stars by \cite{Lee2007}.
Actually, since we find that the polar Alfv\'{e}n modes form a discrete spectrum,
one can safely study the problem by using modal decomposition.
Additionally, \cite{Lee2007} observed that for oscillations confined in the crust there exists a critical
value for the magnetic field strength. That is,  the frequencies of the magnetic modes
become smaller as the magnetic field becomes weaker and then, depending on the stellar model, they seem to disappear after a critical value. We have observed a similar disappearance of the modes, but we cannot give a conclusive answer since this might be a result of the numerical method that we used. Specifically, as the magnetic field becomes weaker the energy on the Alfv\'{e}n modes becomes minimal and the Fourier transform is dominated by the fluid modes, thus it becomes difficult if not impossible to extract the frequencies of the Alfv\'{e}n modes. Thus it is risky to conclude with time evolutions that there is a clear cut off frequency. Instead, in order to answer this, mainly academic, question we plan to study the problem by using modal analysis of the perturbation equations .
Concluding, the observation by \cite{Lee2007}  remains to be proved for the polar Alfv\'{e}n modes in the framework of general relativity.

\begin{figure}
\begin{center}
\begin{tabular}{cc}
\includegraphics[scale=0.42]{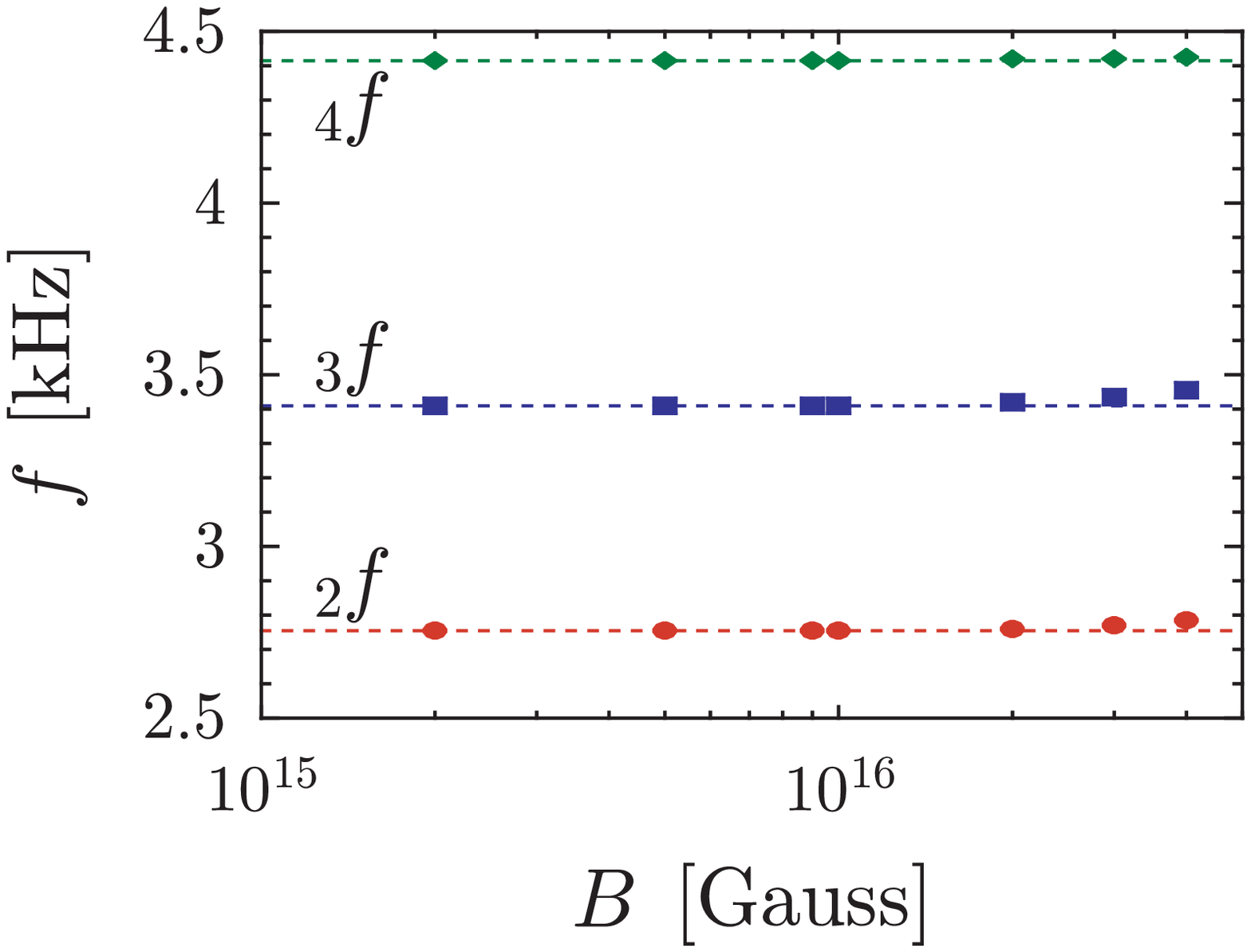} &
\includegraphics[scale=0.42]{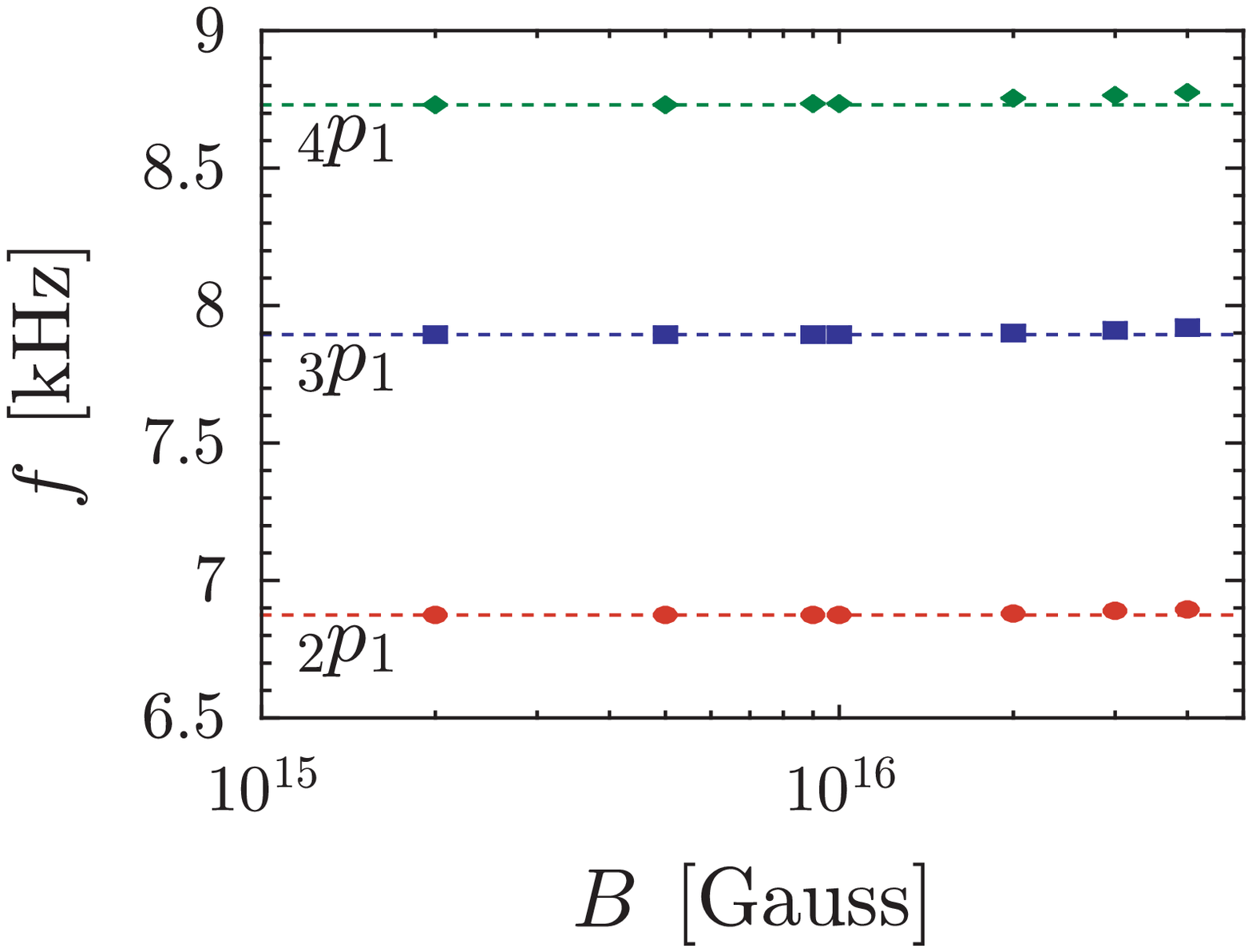} \\
\end{tabular}
\caption{
The frequencies of the $f$ and $p_1$ modes with harmonic indexes $\ell=2$, $3$ and $4$ as functions of
magnetic field strength. Marks denote the corresponding frequencies determined by using
FFT after time evolution, while the horizontal dashed lines show the frequencies
of $f$ and $p_1$ modes without magnetic fields.
}
\label{fig:fluid}
\end{center}
\end{figure}

Finally, we examine how the fluid modes depend on the magnetic field strength.
For a non-rotating magnetars with cold EoS there exists only the family of pressure modes
 ($f$ and $p_i$ modes) 
in addition to the polar Alfv\'{e}n modes mentioned earlier.
However, these fluid modes are hardly affected in any way from the presence of the ultra strong 
 magnetic field unless its strength is larger than about $10^{17}$ Gauss.
This weak dependence on the magnetic field can be understood by realizing that restoring force for these fluid modes is the pressure and the magnetic pressure is typically
 very weak compared to the pressure of the fluid. 
In figure \ref{fig:fluid}, we plot the frequencies of the $f$ and $p_1$ modes with harmonic indexes
$\ell=2$, $3$ and $4$ as functions of magnetic field strength.
The marks correspond to  frequencies of the modes of the magnetized stars
while the horizontal dashed lines are the frequencies for the non-magnetized ones.
From this figure, we can see that the effect of magnetic field on the frequencies
of fluid modes can be traced only when the magnetic field becomes stronger than a few times $10^{16}$ Gauss. Still the deviations
between the frequencies for non-magnetized stars and those for magnetars with $B=4\times 10^{16}$ Gauss
are very small such as 1.1 \% for ${}_2f$, 1.3 \% for ${}_3f$ and 0.3 \% for ${}_4f$ modes
while 0.4 \% for ${}_2p_1$, 0.3 \% for ${}_3p_1$ and 0.7 \% for ${}_4p_1$ modes.
Since the frequencies of the polar Alfv\'{e}n modes depend on the magnetic field it is possible that for ultra strong magnetic fields to reach those of the $f$ modes.

%

\section{Conclusion}
\label{sec:V}

In this paper we study the polar type oscillations of strongly magnetized neutron stars.
As a first step we derive the perturbation equations for an arbitrary magnetic field
although for the numerical studies we adopted a stellar model with a global dipole magnetic field.
In this study we ignored the effects due to the presence of the crust,
since these types of modes are confined mainly in the core of the magnetar.
By using the 2D time evolutions of the perturbation equations,
together with the appropriate initial data sets,
we estimate the Alfv\'{e}n oscillation modes as well as the fluid modes.
We used two techniques to examine the form of the oscillation spectrum,
i.e., whether is continuous as in the case of axial perturbations or discrete.
Actually, by examining the FFT amplitude at various points inside the star and phase of the corresponding frequencies,
we conclude that the Alfv\'{e}n
oscillations with polar parity are described by a spectrum consisting only by discrete modes.
There is a physical reason why the polar oscillations form a discrete spectrum while the axial ones a continuum.
For the axial oscillations  the restoring force is only the magnetic one and  around the magnetic pole the direction of the magnetic force becomes complicated, see also the discussion in \cite{SKS2008,CBK2009,DSF2009}.
On the other hand, for polar oscillations the restoring force is not only the magnetic one
but also the pressure. Thus in the absence of magnetic field these modes are still present while the axial ones reduce to a zero frequency spectrum. As we mentioned earlier, this behavior reminds the spectral properties observed the oscillations of rotating stars. There the axial modes that had as restoring force only the Coriolis one were showing a continuous spectrum (see e.g. \cite{Kojima1998,BK1999}) while the polar ones were admitting a discrete spectrum .

Finally the frequencies of the fundamental polar Alfv\'{e}n oscillations for typical magnetars are of the order of
a few hundred Hz and their dependence  on the magnetic field, can be used to explain 
some of the observed high QPO frequencies in SGRs. 

This is the first study of its kind for axisymmetric polar Alfv\'{e}n oscillations,
it remains open to understand the form of the spectrum for non-axisymmetric perturbations
while more complicated forms of the magnetic field geometry should be assumed.
For example, the inclusion of a toroidal component as in \cite{SCK2008} will
affect the form of the oscillation spectrum. Finally, the inclusion of the crust
enriches the spectrum (\cite{VKS2008}) and the various new features have to be
taken into account in the attempts to understand the structure of magnetars via their QPOs.
Actually, the introduction of crust, although it makes the problem more complicated might erase partially or completely the degeneracy in axial mode spectrum.

\section*{Acknowledgments}

We thank Erich Gaertig for valuable comments.
This work was supported via the Transregio 7 ``Gravitational Wave Astronomy"
financed by the Deutsche Forschungsgemeinschaft DFG (German Research Foundation).



\end{document}